\documentclass[
 reprint,
 superscriptaddress,
 amsmath,amssymb,
aps,
pra,
floatfix,
]{revtex4-1}

\usepackage{graphicx}
\usepackage{xcolor}
\usepackage{amsmath}
\usepackage{soul}
\usepackage{multirow}

\def\bk{\boldsymbol{k}}

\begin{document}

\title{Multifold topological semimetals}

\author{Iñigo Robredo}
\affiliation{Max Planck Institute for Chemical Physics of Solids, 01187, Dresden, Germany}
\affiliation{Donostia International Physics Center, 20018, Donostia - San Sebastian, Spain}

\author{Niels Schröter}
\affiliation{Max Planck Institut f\"ur Mikrostrukturphysik, 06120, Halle, Germany}

\author{Claudia Felser}
\affiliation{Max Planck Institute for Chemical Physics of Solids, 01187, Dresden, Germany}

\author{Jennifer Cano}
\affiliation{Department of Physics and Astronomy, Stony Brook University, 11794, Stony Brook, USA}
\affiliation{Center for Computational Quantum Physics, Flatiron Institute, 10010, New York, USA}

\author{Barry Bradlyn}
\affiliation{Department of Physics, University of Illinois, 61820, Urbana-Champaign, USA}

\author{Maia G. Vergniory}
\affiliation{Max Planck Institute for Chemical Physics of Solids, 01187, Dresden, Germany}
\affiliation{Donostia International Physics Center, 20018, Donostia - San Sebastian, Spain}

\begin{abstract}
The discovery of topological semimetals with multifold band crossings has opened up a new and exciting frontier in the field of topological physics. These materials exhibit large Chern numbers, leading to long double Fermi arcs on their surfaces, which are protected by either crystal symmetries or topological order. The impact of these multifold crossings extends beyond surface science, as they are not constrained by the Poincaré classification of quasiparticles and only need to respect the crystal symmetry of one of the 1651 magnetic space groups. Consequently, we observe the emergence of free fermionic excitations in solid-state systems that have no high-energy counterparts, protected by non-symmorphic symmetries. In this work, we review the recent theoretical and experimental progress made in the field of multifold topological semimetals. We begin with the theoretical prediction of the so-called multifold fermions and discuss the subsequent discoveries of chiral and magnetic topological semimetals. Several experiments that have realized chiral semimetals in spectroscopic measurements are described, and we discuss the future prospects of this field. These exciting developments have the potential to deepen our understanding of the fundamental properties of quantum matter and inspire new technological applications in the future.
\end{abstract}

\maketitle

\section{Introduction}\label{intro}

The prediction of Dirac and Weyl fermions dates back to 1928 \cite{Dirac} and 1929 \cite{Weyl} respectively, when Dirac proposed a four-component wave function to describe massive spin-1/2 fermions and Weyl simplified it to a two-component spinor to describe massless particles. Despite decades of searching by particle physicists, Weyl fermions remained elusive for almost a century. However, in recent years, the search has shifted to condensed matter systems, where Weyl and Dirac fermions are observed as protected crossings in electronic Bloch bands near the Fermi level, in two (Dirac) and three dimensions (Dirac and Weyl). This shift in research has opened up new avenues for investigating the properties and applications of these exotic particles.       

The first attempt to find a Dirac fermion in a solid state system was made in 1987, when Boyanovsky {\it et al.} \cite{Boyanovsky_1987} proposed PbTe as a platform to realize the parity anomaly of 2 + 1 dimensional QED, after the prediction of a Dirac crossing in its band structure. Unfortunately, the model was incomplete and the theoretical prediction proved incorrect \cite{PbTe}. A successful  identification of a solid state Dirac fermion was delayed until many years later in graphene \cite{RevModPhys2009}, a two-dimensional (2D) material consisting of a single layer of carbon atoms arranged in a hexagonal lattice. The two low-energy electronic bands in graphene are doubly degenerate and meet at two linearly dispersing crossing points, known as the Dirac cones or Dirac points because near these points electrons behave as  massless Dirac fermions. Later, the 3D generalizations of graphene, known as Dirac semimetals, were predicted \cite{Wang2013,Wang2012} and observed \cite{YulinNatMat2014,YulinScience2014} in Cd$_3$As$_2$ and Na$_3$Bi.

It took longer to discover Weyl fermions in solid state systems. The first prediction of the ubiquity of perturbatively stable, ungappable 2-fold crossings in noncentrosymmetric or non-TR-invariant crystals (Weyl nodes) was made by Herring in 1937 \cite{Herring}. The first proposal on a realistic material, though, came much later. They were predicted in pyrochlore iridates \cite{Wan_2011,yang2011quantum}, which was followed by a roadmap to engineer Weyl fermions in topological insulator heterostructures~\cite{BurkovWeyl2011}. Unlike Dirac fermions, Weyl fermions are unique to 3D systems, and they exhibit twofold band crossings near the Fermi level with linear dispersion in all three dimensions. They are protected by a topological invariant known as the Chern number, which allows them to appear at any point within the Brillouin zone (BZ). Soon after, the first manifestation of a Weyl node in a crystal was pointed out theoretically in HgCr$_2$Se$_4$ by Wan {\it et al}\cite{Xu2011ChernSA} and a Weyl node in Hg$_{1 - x - y}$Cd$_x$Mn$_y$ by Bulmash {\it et al}\cite{Bulmash}. Unfortunately, none of these systems came to fruition experimentally. Weyl semimetals became finally a reality with the experimental discovery \cite{DingPRX2015,SYXuScience2015} confirming  the theoretically predicted \cite{XDaiPRX2014,SMHuangNatComm2015} Weyl nodes in the TaAs family of compounds.  However, finding new materials exhibiting Weyl fermions remained a challenge.

A big step forward was taken when it was realized that the co-representations (coreps) of the little group \cite{BC-book} at the high-symmetry points in the BZ could correspond to nodal fermions \cite{manes2012,Young2012}. This discovery set the stage for the machinery to systematically identify new materials using first-principles calculations combined with crystal symmetry and group theory \cite{Steinberg2014,Wieder2016}. In 2016 Bradlyn {\it et al} \cite{NewFermions} and Wieder {\it et al} \cite{Wieder8fold}  recognized that non-symmorphic crystalline symmetries could stabilize previously undiscovered crossings with no analogs in high-energy physics. In these systems, the low-energy $\bk\cdot\boldsymbol{p}$ Hamiltonians do not resemble the dispersion relations of any relativistic particle. Yet, they exhibit an isotropic limit resembling a Weyl fermion, $H=\bk\cdot \boldsymbol{S}$, with $S_i$ a set of $n$-dimensional spin matrices. Initially dubbed the multifold fermions, these excitations introduced the field of multifold topological semimetals. The newly introduced band crossings can be 3-, 4-, 6- and 8-fold degenerate at the crossing point, which may or may not exhibit a topological charge depending on their crystal symmetry. In current parlance, these $n$-dimensional fermions are referred to as \emph{$n$-folds}, with \emph{n} denoting the dimension of the band crossing, which is the notation we will use throughout this work.
This work reviews the classification of these n-fold fermions, along with recent theoretical and experimental advances.

The manuscript is organized in the following way: in Section \ref{classification} we will give a full classification of high-order crossings in the 1651 Shubnikov groups and provide a general procedure to identify these nodes in materials. Section \ref{experiments} will review the recent progress in materials realization with particular focus in \textit{chiral} crystals. In Section \ref{sec:3} we provide the computational details of the band structure calculations in Section~\ref{experiments}. We conclude the manuscript with a discussion of future perspectives.

\section{Classification of all Shubnikov groups and general procedure}\label{classification}

In this section, we classify the symmetry-protected nodal points that can exist at high symmetry points within magnetic space groups by first reviewing the Shubnikov groups that describe crystalline materials and then discussing the dimensionality of their coreps.

\subsection{Magnetic Space Group Types}

Multifold fermions, unlike Weyl fermions, need symmetry protection to be stabilised. In a particular symmetry group, the symmetry operations and their commutation relations dictate the multifold fermions that the group can protect. We thus start by reviewing the classification of Shubnikov groups, which will serve as a basis for the classification of multifold fermions in crystalline materials.

There are 1651 Shubnikov groups, also named magnetic space groups (MSG), which can be classified into four types. First, the type I groups are those that contain only unitary crystalline symmetries, i.e., no time reversal symmetry (TRS) or combinations of crystalline symmetries and TRS. There are 230 such groups. They can describe some materials with magnetic order, but only a reduced subset of all commensurate magnetic orderings. Nonetheless, the type I groups provide the underlying structure from which the remaining MSGs can be constructed. Starting from a type I MSG, $H$, adding TRS as a group element creates a type II MSG (non-magnetic, often referred to as `gray' groups), $G=\{E,\mathcal{T}\}\times H=H\,\cup \mathcal{T} H$, with $\mathcal{T}$ the TRS operator. These groups describe non-magnetic crystals, such as the ones analysed in the topological quantum chemistry database \cite{TQC_databse,Alltopo,Catalogue_Zhang,Tang2019}. The remaining two types (type III and IV), the so called black-white groups, deal with the combination of TRS and unitary symmetries and can describe the remaining commensurate magnetic orderings. These MSGs can be written as $G=H\cup \mathcal{T} g_0 H$, with $H$ a type I MSG and $g_0 \notin H$ a unitary symmetry. In this context, type IV groups are those that contain fractional translations in $g_0 H$, while type III MSGs do not.

This exhausts the classification of magnetic symmetry groups, which can be used to describe all commensurate magnetic orderings. A further sub-classification of MSGs is obtained by distinguishing between the presence or absence of improper symmetries, that is, symmetries with determinant $-1$. If a MSG does not contain elements such as rotoinversions or mirrors that reverse spatial orientation it is called a Sohnke MSG and it gives rise to a \emph{structurally chiral} crystal, while if it contains inversion symmetry, rotoinversions or mirrors it is non-chiral. This distinction is relevant because \emph{only Sohnke MSGs can host topologically charged multifold fermions}. In non-Sohnke MSGs, symmetry prevents a multifold fermion from carrying a topological charge.

We now turn to identifying the MSGs that can host $n$-fold crossings in the presence of spin-orbit coupling (SOC).
The first step is to find the $n$-dimensional ($n = 3, 4, 6, 8$) irreducible coreps of each of the 1651 MSGs: an irreducible corep corresponds to an $n$-dimensional symmetry enforced degeneracy. This search can now be easily carried out by using the Bilbao Crystallographic Server (BCS) \cite{BCS1,BCS2,BCS3}. Preliminary work on this classification was carried out by Bradlyn et al \cite{BradlynEA17}, Chang et al \cite{chang2018topological} and Wieder et al \cite{Wieder8fold} for type II MSGs and generalized to all MSGs by Cano et al \cite{cano2019multifold}.

\subsection{Multifold Fermions}

\begin{table*}
\caption{List of MSGs containing 3-fold, 6-fold and 8-fold fermions in systems with significant SOC. Each cell lists the MSG and the $\mathbf{k}$-point at which the multifold fermion resides.}
    \label{tab:multifolds}
\begin{center}
    \centering
    \begin{tabular}{c|cc|cc|cc|cc}
          & \multicolumn{2}{c|}{Type I}  & \multicolumn{2}{c|}{Type II} & \multicolumn{2}{c|}{Type III}  & \multicolumn{2}{c}{Type IV} \\\hline
        \multirow{4}{*}{3-fold}   & \scriptsize{199.12, 214.67, 220.89} & \multirow{2}{*}{\scriptsize{P}} & \scriptsize{199.13, 214.68, 220.90} & \scriptsize{P} & \scriptsize{214.69, 220.91, 230.148} & \scriptsize{P} &  &  \\
        & \scriptsize{206.37, 230.145} &  &&&&&  &  \\
           & \scriptsize{198.9, 205.33, 212.59} & \multirow{2}{*}{\scriptsize{R}}&  &  & \scriptsize{212.61, 213.65} & \scriptsize{R} & \scriptsize{198.11, 212.62, 213.66} & \scriptsize{R}\\
           & \scriptsize{213.63} &  &&&&&  &  \\\hline
        \multirow{2}{*}{6-fold} &  &  &  \scriptsize{206.37, 230.145} & \scriptsize{P} & \scriptsize{206.39, 230.147, 230.149} & \scriptsize{P} &  & \\
           &  &  &  \scriptsize{198.10, 205.34, 212.60, 213.64} & \scriptsize{R}  & \scriptsize{205.35} & \scriptsize{R} & \scriptsize{205.36} & \scriptsize{R} \\\hline
          \multirow{4}{*}{8-fold} &  &  & \scriptsize{130.424, 135.484} & \scriptsize{A} &  &  & \scriptsize{125.374, 126.385, 129.420} & \multirow{2}{*}{\scriptsize{A}}\\
           &  &  &  &  &  &  & \scriptsize{131.445, 132.458, 136.504} & \\
          &  &  & \scriptsize{218.82, 222.99, 223.105} & \scriptsize{R} & \scriptsize{222.102, 223.108} & \scriptsize{R} & \scriptsize{215.73, 221.97, 224.115} & \scriptsize{R}\\
          &  &  & \scriptsize{220.90, 230.146} & \scriptsize{H} & \scriptsize{230.149} & \scriptsize{H} &  & \\
    \end{tabular}
    \end{center}
\end{table*}

\begin{figure*}[t]
    \centering
    \includegraphics[width=\linewidth]{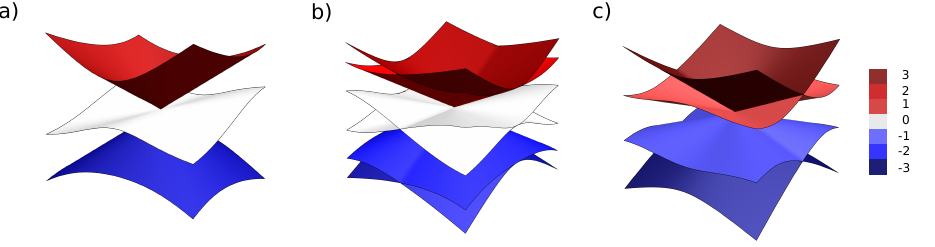}
    \caption{Topologically charged multifold fermion dispersions. a) 3-fold spin-1, b) 6-fold double spin-1 and c) 4-fold RS fermion.
    The color code shows the topological charge (Chern number) of each band.}
    \label{fig:multifolds}
\end{figure*}

We start by analysing the 3-fold fermion, which was first predicted in the type I MSG I$2_131'$ (199.13) \cite{BradlynEA17} as the $\bk\cdot\boldsymbol{p}$ Hamiltonian in the vicinity of the $P=(1/4,1/4,1/4)$ point. This turns out to be the most general 3-fold fermion, to which all 3-fold fermions in all MSGs are equivalent. The continuum Hamiltonian reads:

\begin{equation}\label{eq:3fold}
    H_{\text{3-fold}}(\phi,\bk)=
    \begin{pmatrix}
    0 & e^{i\phi}k_x & e^{-i\phi}k_y \\
    e^{-i\phi}k_x & 0 & e^{i\phi}k_z \\
    e^{i\phi}k_y & e^{-i\phi}k_z & 0
    \end{pmatrix},
\end{equation}
where $\phi$ is a free parameter in the type II MSGs but can be symmetry-restricted in other MSGs. In the particular case of $\phi=\pi/2$, the Hamiltonian in Eq.~\ref{eq:3fold} takes the familiar form $H_{\text{3-fold}}(\pi/2,\bk)=\bk\cdot\boldsymbol{S}$, with $\boldsymbol{S}_{x,y,z}$ the generators of the spin $s=1$ representation of $SU(2)$. Notice that it resembles the Weyl Hamiltonian, $H_{\text{Weyl}}=\bk\cdot\boldsymbol{\sigma}$ (with $\boldsymbol{\sigma}$ the vector of Pauli matrices), which is the special case of $s=1/2$. Hence the 3-fold fermions are referred to as spin-1 Weyl fermions. The $H_{\text{3-fold}}$ Hamiltonian is gapped away from the degeneracy point, so the Chern number of each band can be computed by computing the Chern number of a sphere enclosing the degeneracy point. This calculation reveals that the Chern number is $-2$/$+2$ for the lower/upper bands and 0 for the middle band (see Fig. \ref{fig:multifolds}a). Noticeably, the 3-folds in type IV MSGs have an additional symmetry generator $g=\{E|\frac{1}{2}\frac{1}{2}\frac{1}{2}\}\circ\mathcal{T}$, which forces $\phi=\pi/2$. This Hamiltonian, which can only be achieved by fine-tuning in type I/II/III MSGs, provides an ideal situation where the energy gap away from the degeneracy point is maximized, thus allowing maximum range for Fermi arcs \cite{cano2019multifold}. In Table~\ref{tab:multifolds} we list all MSGs which can host 3-fold fermions in systems with significant SOC and at which high symmetry $k$-point.

The 6-fold fermions can be constructed by using the 3-fold fermion as a building block. There are two classes of 6-fold fermions depending on whether the MSG is chiral: double spin-1 fermions and spin-1 Diracs, respectively. In chiral type II MSGs, TRS pins two symmetry-related copies of the 3-fold fermion in the same $k$-point. The bands are gapped in generic directions away from the degeneracy point and the topological charge distribution can be seen in Fig. \ref{fig:multifolds}b. The two lower bands have topological charges of -2 each, while the middle ones have 0 charge, which amounts to a total of -4 at half-filling, the maximum topological charge of multifold fermions in the MSGs. In non-chiral type II MSGs, inversion (or rotoinversion) symmetry pins two 3-folds with opposite topological charge at the same $k$-point. In this case, the bands are doubly degenerate in generic directions away from the degeneracy point and the topological charge is 0 for all bands. These 6-folds are usually called spin-1 Dirac fermions, since the original Dirac fermion can be decomposed into two spin-1/2 Weyl fermions of opposite charge. In type III MSGs, all 6-folds are of spin-1 Dirac type, due to the presence of $\mathcal{I}\circ\mathcal{T}$ symmetry. In type IV MSG P${}_Ia\bar{3}$ (205.36), the 6-fold is also spin-1 Dirac, but the extra symmetries in the little co-group force the parameters to be pinned at an exactly solvable point in parameter space, similarly to what happens to the 3-folds in type IV MSGs.

It is natural to ask whether there are higher-dimensional effective Hamiltonians of the form $H=\bk\cdot\boldsymbol{S}$. An exhaustive search of all MSGs revealed a 4-dimensional version of this Hamiltonian, usually referred to as a Rarita-Schwinger (RS) or spin-3/2 fermion, due to its analogy to high-energy physics\footnote{Technically, the RS fermion Hamiltonian is not analytic in $k$ (Eq.~(7) in Ref.~\cite{cano2019multifold}) because the second term contains $k^2$ in the denominator. The authors decided to rename only the analytic part as RS$^*$ fermion, which is the Hamiltonian of the form $H_{3/2}=\bk\cdot\boldsymbol{S^{3/2}}$ with $S^{3/2}_{x,y,z}$ the generators of the spin $s=3/2$ representation.} \cite{cano2019multifold,liang2016semimetal,tang2017multiple}. In this case, the $S_{x,y,z}$ matrices are the generators of the spin $s=3/2$ representation (also denoted as $S^{3/2}_{x,y,z}$) and the resulting Chern numbers of the bands are $-3$, $-1$, $1$ and $3$, respectively (see Fig. \ref{fig:multifolds}c). At half filling (two occupied bands) these quasiparticles have maximum topological charge, $1+3=4$.
Noticeably, they can only be hosted in Sohnke MSGs, as rotoinversions pin two copies with opposite charge at the same point. In Ref.~\cite{BradlynEA17} the authors list the type II MSGs where these multifolds can occur. However, there has been no extensive search for spin-3/2 multifold fermions in type III or type IV MSGs.

Analogous to the 3-fold, for certain material parameters the 4-fold Hamiltonian is adiabatically connected to the $H_{3/2}=\bk\cdot\boldsymbol{S^{3/2}}$ Hamiltonian described before, with Chern numbers -3,-1,1,3. This phase is realized in most known materials with fourfold fermions (see Sec.~\ref{experiments}. However, the 4-fold $\boldsymbol{k}\cdot\boldsymbol{p}$ Hamiltonian also admits phases not adiabatically connected to $H_{3/2}$ \cite{BradlynEA17}. It was recently shown~\cite{Schnyder_other_multifold} that for certain parameter values the fourfold can exhibit Chern numbers $-3, +5, -5, +3$ in the four bands. In the same work, they propose BaAsPt as a material displaying this phase of the 4-fold fermion. Finding experimental prove of a material that realizes this interesting phase is an active area of study.

Similar to how (spin-1/2 or spin-1) Dirac fermions can be decomposed into two (spin-1/2 or spin-1) Weyl fermions with opposite chirality, which are pinned to the same energy and momentum by symmetry, certain 8-fold fermions can be decomposed into two copies of 4-folds. The 8-fold degeneracy is the highest possible in crystalline solids \cite{TQC}. The 8-fold fermions hosted in cubic MSGs are formed by two spin-3/2 4-fold fermions with opposite charge that are pinned together by symmetry. The result is doubly degenerate bands in generic directions and no net topological charge. 

The 8-fold fermions in the tetragonal MSGs, though, are fundamentally different. 
In that case, if inversion symmetry is removed as a group generator, a 4-dimensional corep yields a fermion isomorphic to a spin-1/2 Dirac. Inversion symmetry doubles this corep to create the 8-fold fermion. Thus, the tetragonal 8-fold can be thought of as a double spin-1/2 Dirac. 

These two types of 8-fold fermions -- the doubled spin-3/2 fermion and doubled spin-1/2 Dirac -- are the only types of 8-fold fermions protected by crystalline symmetry. The two of them have in common that all bands are doubly degenerate in arbitrary directions from the node and have 0 topological charge.

We now turn to the type III and type IV MSGs.
In type III MSGs the same results apply, since a type III MSG with an 8-fold degeneracy is a subgroup of a type II MSG with that same 8-fold degeneracy. In type IV MSGs, all 8-folds are of the double spin-1/2 Dirac form. However, similar to the 3-fold and 6-fold fermions in type IV MSGs, the little cogroup symmetries pin the Hamiltonian to a particularly simple form. In this case, the result is 8-fold fermions that are perfectly isotropic double spin-1/2 Dirac fermions. In Table \ref{tab:multifolds} we show all MSGs that can host 8-fold fermions.

\section{Material realizations}\label{experiments}

\begin{figure*}[t]
    \centering
    \includegraphics[width=\linewidth]{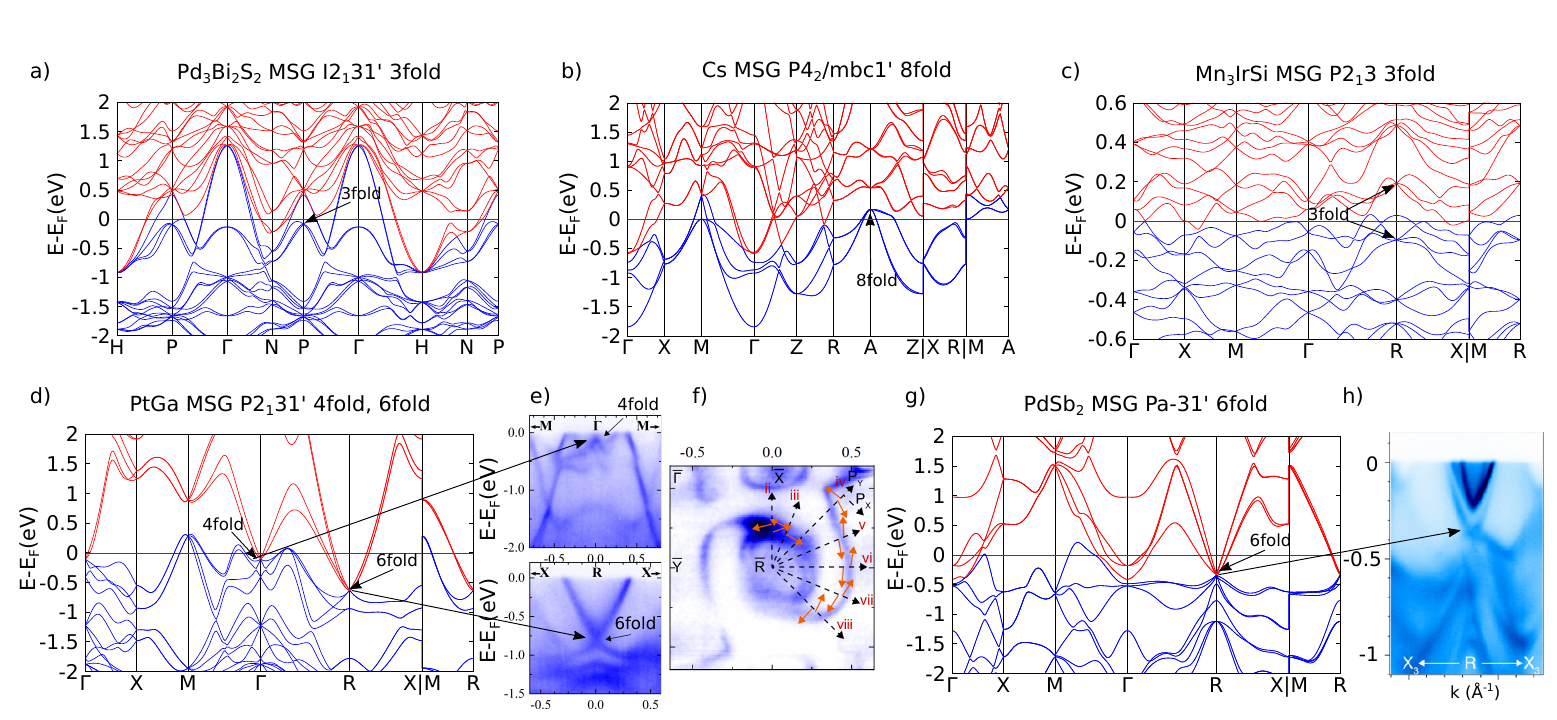}
    \caption{Crystal structures and electronic bands of non-magnetic and magnetic materials displaying multifold fermions. a) Pd${}_3$Bi${}_2$S${}_2$, a chiral material displaying a 3-fold fermion at the P point \cite{Alltopo,TQC_databse}. b) Elemental Cs, in non-Sohnke MSG P$4_2$/mbc1', displaying an 8-fold fermion close to the Fermi level at A. c) Mn${}_3$IrSi in MSG P2${}_1$3 with chiral magnetic ordering displays 3-fold spin-1 fermions at R. d) PtGa in Sohnke MSG P2${}_1$31', displaying the maximum charge RS 4-fold at $\Gamma$ and double spin-1 6-fold at R. e) ARPES measurements reported in Ref.~\cite{PtGa_exp} showing the bulk band dispersion of both multifold fermions. f) Surface ARPES measurements showing Fermi arcs and their spin polarization. g) PdSb${}_2$ in non-Sohnke MSG Pa-31' presenting a spin-1 Dirac 6-fold fermion at R. h) ARPES measurements reported in Ref.~\cite{6fold_Dirac_exp_PdSb2} showing the bulk dispersion of the spin-1 Dirac 6-fold (extracted from Ref.~\cite{6fold_Dirac_exp_PdSb2}). Notice that ARPES measurements for PdSb${}_2$ were performed at the VUV range \cite{6fold_Dirac_exp_PdSb2}, while they were performed at the SX range for PtGa \cite{PtGa_exp}.}
    \label{fig:bandplots}
\end{figure*}

In this section we will describe the material realisations of the multifolds described in previous sections.

\subsection{Non-magnetic materials}

The theoretical prediction of multifold fermions led to a huge advance in material realisations, particularly in nonmagnetic systems. 3-fold fermions have been reported in chiral Pd${}_3$Bi${}_2$S${}_2$ in SG I$2_13$1' (199.13) \cite{Alltopo,TQC_databse} (see Fig. \ref{fig:bandplots}a). 4-fold and 6-fold double-spin-1 fermions can be found in the B20 material family in the presence of SOC. Materials such as the previously mentioned (Rh,Co)(Si,Ge) \cite{TMSi_multis}, AlPt \cite{AlPt}, PdGa \cite{PdGa} and PtGa \cite{PtGa} (see Fig. \ref{fig:bandplots}d) are excellent examples of materials hosting these fermions, with extensive theoretical predictions and experimental confirmation of topological properties such as very long Fermi arcs connecting the surface projections of the multifold fermions. 6-fold Dirac fermions have been reported in non-chiral PdSb${}_2$ \cite{6fold_Dirac_exp_PdSb2} (see Fig.~\ref{fig:bandplots}g and Fig.~\ref{fig:bandplots}h), PtBi${}_2$ \cite{6fold_Dirac_exp_PtBi2}, Rb${}_4$O${}_6$ \cite{6fold_Dirac_theo_Rb4O6}, Li${}_12$Mg${}_3$Si${}_4$ \cite{6fold_Dirac_theo_Li12Mg3Si4} and SrGePt, which also displays a 4-fold fermion \cite{multiSrGePt}.

Finally, 8-fold fermions have also been predicted and found in non-magnetic materials. 
The 8-fold double-Dirac fermions can exist in seven type II SGs and were predicted in Bi${}_2$AuO${}_5$ in MSG P4/ncc1' (130.424)~\cite{8fold_theo_Bi2AuO5}. Elemental Cs crystals in MSG P$4_2$/mbc1' (135.484) also host an 8-fold crossing close to the Fermi level \cite{TQC_databse,Alltopo} (see Fig. \ref{fig:bandplots}b).

\subsection{Magnetic materials}

Multifolds in magnetic materials are inherently harder to find, due to the relative scarcity of high quality magnetic materials' crystal and magnetic structures reported in the literature. 

The most notable example is that of the Mn${}_3$IrSi family (Mn${}_3$IrGe, Mn${}_3$ Ir${}_{1-y}$Co${}_y$Si, and Mn${}_3$CoSi${}_{1-x}$ Ge${}_x$) with type I MSG P$2_13$ (198.9), which displays a rare example of chiral magnetism. It was predicted to host 3-fold fermions at the R point below the Néel temperature \cite{cano2019multifold} (see Fig. \ref{fig:bandplots}c). Above the critical temperature, the magnetic order is destroyed and we recover the full gray group P$2_13$1' (198.10), in which the 3-folds are forced to occur in pairs due to TRS.

Following recent advances in the development of high-quality databases for magnetic materials (for instance, MAGNDATA \cite{MAGNDATA}), we are now able to predict several materials in type III MSGs that can host 3-fold fermions, such as BaCuTe${}_2$O${}_6$ in MSG P$4_1'32'$ (213.65) at the $R$ point \cite{3fold_TypeIII} and the family of La${}_3$X${}_5$O${}_12$ (where La is a lanthanide and X=Ga,Al) in MSG I$a\bar{3}d'$ (230.148) at the $P$ point \cite{3fold_TypeIII_2,3fold_TypeIII_3,3fold_TypeIII_4,3fold_TypeIII_5}. 6-fold fermions can be found in the TRh (T=Dy, Ho, Er) family of compounds in MSG $P_C2_1/m$ (11.57)\cite{6fold_TypeIV}, as well as in Eu${}_3$PbO  with MSG $P_Ia\bar{3}$ (205.36)\cite{6fold_TypeIV_2}, both type IV MSGs. Finally, we predict 8-fold fermions in BaNd${}_2$ZnS${}_5$, with MSG $P_C4/nnc$ (126.385) \cite{8fold_TypeIV}.

\subsection{Experimental evidence}

There are two fundamental experimental signatures of multifold fermions: their bulk band dispersion and Fermi arc surface states. Both can be measured by means of angle-resolved photoemission spectroscopy (ARPES). Due to the short inelastic mean free path photoelectrons in solids of $\lambda\approx 0.5 \text{nm}$ when excited with extreme ultraviolet light (photon energies of $\approx$ 10-150 eV), this technique has been very successful in imaging topological surface states \cite{ARPES1,ARPES2}. In particular, it has been used to visualize the predicted long Fermi arcs connecting the projection of the 4-fold and 6-fold fermions in the B20 material family \cite{PtGa,PtGa_exp,RhSi_exp,RhSi_multi,CoSi_exp} (see Fig.~\ref{fig:bandplots}f). These multifold fermions were predicted to carry the maximum topological charge of four, requiring four Fermi arcs to emerge from the surface projection of the multifolds. The four Fermi arcs can be observed in  materials where the SOC is large enough to lead to a sizable lifting of the spin-degeneracy, for example, close to the $\Gamma$ point in PtGa \cite{PtGa_exp} or in PdGa \cite{PdGa_exp}.
When SOC is small, it is difficult to resolve the spin-splitting, as in CoSi \cite{CoSi_exp}.

To study the bulk electronic structure of solids with ARPES, higher photon energies in the soft X-ray (SX) regime (approx. 150-1500 eV) need to be employed. Here, the inelastic mean free path can typically extend to multiple unit cells, which reduces the effect of momentum broadening in the out of plane direction ~\cite{strocov2003intrinsic}. Soft X-ray ARPES is therefore more suitable to extract information about the three-dimensional bulk band structure. An example of this can be seen in Fig.~\ref{fig:bandplots}e, which shows the bulk dispersion of 4-fold and 6-fold multifold fermions in PtGa \cite{PtGa_exp} in ARPES and as predicted by DFT (Fig.~\ref{fig:bandplots}d), as well as in Fig.~\ref{fig:bandplots}h, which shows the bulk bands of the 6-fold Dirac in PdSb${}_2$. Notice that for the visualization of the sixfold in PdSb${}_2$ in Fig.4\ref{fig:bandplots}, lower photon energies in the Vacuum-ultraviolet (VUV) range were employed \cite{6fold_Dirac_exp_PdSb2}.

Even though most of the experimental research has been focused on materials in the B20 family \cite{RhSi_exp,CoSi_exp,PdGa_exp,RhSi_exp_2,CoSi_exp_2,CoSi_exp_3,CoSi_exp_4,CoSi_exp_5,CoSi_exp_6,FeSi_exp,RhSi_exp_3,CoSiRhSi_exp}, PdBiSe, also in MSG P$2_13$1' (198.10), exhibits 4-fold and 6-fold fermions which have been experimentally observed \cite{PdBiSe_exp}. In non-chiral crystals, 6-fold Dirac fermions have been recently found in cubic SG Pa$\bar{3}$1' (205.34) in PdSb${}_2$ \cite{6fold_Dirac_exp_PdSb2,6fold_Dirac_exp_PdSb2_2,6fold_Dirac_exp_PdSb2_3} and in PtBi2 \cite{6fold_Dirac_exp_PtBi2}. Finally, `practical' 8-fold fermions have been reported in TaCo${}_2$Te${}_2$\cite{8fold_exp_TaCo2Te2} in MSG Pnma1' (62.442) without SOC.

\section{Computational methods}\label{sec:3}

The calculations presented in Figure \ref{fig:bandplots} have been computed for the purpose of this review based on the structural parameters reported in their respective publications \cite{Alltopo,AlPt,6fold_Dirac_exp_PdSb2,TQC_databse,cano2019multifold}. We performed density functional theory (DFT) calculations as implemented in the Vienna Ab Initio Simulation Package (VASP) \cite{VASP1,VASP2,VASP3,VASP4}. The interaction between the ion cores and valence electrons was treated by the projector augmented wave (PAW) method \cite{PAW}, the generalized gradient approximation (GGA) was employed for the exchange-correlation potential with the Perdew–Burke–Ernzerhof for solid parameterization \cite{PBE} and the SOC was considered on the second variation method \cite{DFT-SOC}. We used a $\Gamma$-centered Monkhorst-Pack of $9\times9\times9$ k-point grid for reciprocal space integration and we used a 500eV energy cutoff for plane wave expansion. The electronic ground state was self-consistently converged with an accuracy of $10^{-5}$ eV/unit cell.

\section{Conclusion}

This work has reviewed the recent progress in material realizations of multifold fermions from a three-pronged perspective, namely, symmetry arguments, DFT predictions of realistic materials, and experimental realizations. In addition to the basic bulk and surface properties described here, multifold fermions have been predicted to display other interesting physical responses, such as second harmonic generation in B20 RhSi \cite{Second_Harmonic_RhSi}, quantized circular photogalvanic effect in multifold fermions \cite{CPGE}, non-linear Hall conductivity in CoSi \cite{Nonlinear_Hall_CoSi} and a new hydrodynamic responses in the B20 family \cite{3fold_Hallvisco}. There has also been research on unconventional superconductivity, with an unusual s-wave triplet pairing proposed in topological semimetals hosting 3-fold fermions \cite{Spin1_superconduct}. Finally, the robustness of multifold fermions to disorder is an important question that deserves future study, as Fermi arcs from 3-fold fermions retain their sharpness at weak disorder \cite{3fold_disorder}.

With the rapid growth of material predictions in recent years, we expect to see more experimental realizations of these exotic multifold fermions in the future. In particular magnetic systems have recently been the focus of many high-throughput searches \cite{MagneticHT,HighThrougput_collinear}. These new materials will prove to be excellent platforms to study multifold fermions and their phenomenology.
\\

\section{Acknowledgements}
M.G.V. and C.F. thank support from the Deutsche Forschungsgemeinschaft (DFG, German Research Foundation) GA3314/1-1 -FOR 5249 (QUAST). M.G.V and I.R. thank support to the Spanish Ministerio de Ciencia e Innovacion grant PID2022-142008NB-I00 and the
Ministry for Digital Transformation and of Civil Service of the Spanish Government through the QUANTUM ENIA project call - Quantum Spain project, and by the European Union through the Recovery, Transformation and Resilience Plan - NextGenerationEU within the framework of the Digital Spain 2026 Agenda. This project was partially supported by the European Research Council (ERC) under the European Union’s Horizon 2020 Research and Innovation Programme (Grant Agreement No. 101020833). B.B. acknowledges the support of the Alfred P. Sloan foundation, and the National Science Foundation under grant DMR-1945058. J.C. acknowledges the support of the Alfred P. Sloan foundation, the National Science Foundation under grant DMR-1942447, and the Flatiron Institute, a division of the Simons Foundation. N.B.M.S. was funded by the European Union (ERC Starting Grant ChiralTopMat, project number 101117424). Views and opinions expressed are however those of the author(s) only and do not necessarily reflect those of the European Union or the European Research Council Executive Agency. Neither the European Union nor the granting authority can be held responsible for them. C.F. was financially supported by Deutsche Forschungsgemeinschaft (DFG) under SFB1143 (Project No. 247310070) and Würzburg-Dresden Cluster of Excellence on Complexity and Topology in Quantum Matter—ct.qmat (EXC 2147, project no. 390858490).

\bibliographystyle{unsrt}
\bibliography{main}

\end{document}